\documentclass[prl,showpacs,preprintnumbers,amsmath,amssymb]{revtex4}
\usepackage{amsfonts}

\usepackage{graphicx}
\usepackage{dcolumn}
\usepackage{bm}
\usepackage{mathrsfs}
\usepackage{graphicx}
\usepackage{float}
\usepackage{amsmath}
\usepackage{subfigure}

\begin{document}

\title{ On six-photon entangled state emitted from a single third-order parametric down-conversion process }

\author{Ying-Qiu He}
\affiliation{College of Science, North China Institute of Science and Technology, Beijing 101601, China}

\author{Dong Ding}
 \email{ dingdong@ncist.edu.cn}
\affiliation{College of Science, North China Institute of Science and Technology, Beijing 101601, China}
\author{Ting Gao}
 \email{gaoting@hebtu.edu.cn}
\affiliation{ School of Mathematical Sciences, Hebei Normal University, Shijiazhuang 050024, China}
\author{Fengli Yan}
 \email{flyan@hebtu.edu.cn} \affiliation{College of Physics, Hebei Normal University, Shijiazhuang 050024, China}

\begin{abstract}
{ We consider six-photon entangled state emitted from a single third-order parametric down-conversion process.
In the regime of weak nonlinearities, we present a symmetry detector which is capable of analyzing the twin-beam six-photon symmetric states.
By cascading the symmetry detectors, as an application, it is shown that one can purify the desired six-photon entangled state from an arbitrary twin-beam six-photon symmetric state.
With linear optics we propose a fruitful scheme for exploring a class of multimode-photon entangled states from third-order parametric down-conversion process.
Furthermore, we provide a method to generate the six-photon polarization entangled Greenberger-Horne-Zeilinger state based on linear optics and weak nonlinearities.
 }
\end{abstract}
 \maketitle

\section{I. Introduction}

Quantum entanglement is one of the most fascinating phenomenons in quantum information processing  \cite{NC2000,Quantum-entanglement}.
Multiphoton entanglement reveals strictly nonclassical quantum correlations \cite{KLM2001,DDI2006,GMR2013,BCPSW2014,SBB2017}.
Nowadays, the nonlinear optical process of parametric down-conversion (PDC) plays an important role in most quantum optical experiments as a standard method of creating entangled-photon pairs \cite{SPDC1970,PDC1990RT,PDC1995,SB2003,Kok2007,Pan2012}.
By using one or more pulse pumped PDC crystals and overlapping pairs of entangled photons, experimentally, the preparation of photonic entanglement of various forms has been proposed \cite{ZHW1997,PDG2001,LZG2007,Pan-10-photon-GHZ-2016}, but involving only a few photons.
Since the overlaps make them fragile, the observation of a larger number of entangled photons is heretofore challenging in multiphoton interferometry.

In PDC process, a pulsed pump field is capable of emitting a series of superposition states containing $2n$ photons \cite{ZZW1995,ORW1999,LHB2001,NOOST2007Science}.
A typical case of four-photon emission leads to superposition of two independent pairs and indistinguishable four-photon entangled state.
De Riedmatten \emph{et al.} \cite{RSMATZG2004} demonstrate experimentally that in the limit where the duration of the pump pulse is much shorter than coherence time of the photons the four photons can be described as an indistinguishable four-photon entangled state; in the opposite limit, the four photons are in two independent pairs.
Note that the higher-order emission of PDC source possesses strongly robust \cite{BRS2003}, since it runs with a single down-conversion crystal and therefore has no additional interferometric alignment.
Although the probabilities of emitting indistinguishable $2n$-photon are usually slower than emitting a pair of entangled photons \cite{KB2000}, exploring multiphoton entangled states from higher-order emissions of down-conversion source may contribute to various quantum information processing \cite{WZ2001,WSKPGW2008,RWZB2009,HDYG2017SR,DHYG2018QIP} because of their essential properties such as robustness and symmetry.

In this paper, we study six-photon entangled state related to third-order emission of the PDC source.
We construct a twin-beam symmetry detector to evolve and pick out six-photon symmetric states by using a beam splitter and weak nonlinearities.
As an interesting application, it is capable of purifying (or preparing) six-photon entangled states $|\psi_3^{-}\rangle$ by cascading the present symmetry detectors.
Based on linear optical elements we propose a scheme to create a class of six-photon entangled states from third-order emission of PDC source.
Also, we present a simple circuit to project these photons onto the six-photon polarization-entangled Greenberger-Horne-Zeilinger (GHZ) state via weak nonlinearities.

\section{II. Symmetry detector and its application}

A simplified Hamiltonian \cite{KB2000} of the nonlinear interaction in PDC process is given by
$H = {i}\kappa({a}^{\dag}_{H}{b}^{\dag}_{V}-{a}^{\dag}_{V}{b}^{\dag}_{H})+\mathrm{H.c.}$,
where $\mathrm{H.c.}$ means Hermitian conjugate, and $\kappa$ is a coupling constant depended on the nonlinearity of the crystal and the intensity of the pump pulse, and ${a}^{\dag}_{x}$ and ${b}^{\dag}_{x}$, $x \in \{H,V\}$, are respectively the creation operators with horizontal or vertical polarization in spatial modes $a$ and $b$.
This leads to the two-mode squeezed state
\begin{equation}\label{psi}
  \left| \Psi\right\rangle =\textrm{e}^{{-\textrm{i}tH}/{\hbar}}|0\rangle = \frac{1}{\cosh^2\tau}\sum_{n=0}^{\infty}\sqrt{n+1}\tanh^n\tau|\psi_n^{-}\rangle,
\end{equation}
where $\tau=\kappa t/\hbar$ is the interaction parameter with $t$ being interaction time, and
\begin{equation}\label{psi n}
|\psi_n^{-}\rangle = \frac{1}{\sqrt{n!(n+1)!}}({a_{H}^{\dag}}{b_{V}^{\dag}} - {a_{V}^{\dag}}{b_{H}^{\dag}})^{n}|0\rangle
\end{equation}
is $2n$-photon entangled state corresponding to $n$th-order emission with the mean photon number $\langle n\rangle=2\mathrm{sinh}^2\tau$.
For a $2n$-photon entangled state $|\psi_n^{-}\rangle$, note that the photon-state remains unchanged when the photons passing through a symmetric $50:50$ beam splitter (BS).
We turn next to the symmetry detector and its application for a class of twin-beam six-photon entangled state by using linear optics and weak nonlinearities.

\begin{figure}[htbp]
\centerline{ \includegraphics[width=2.8in]{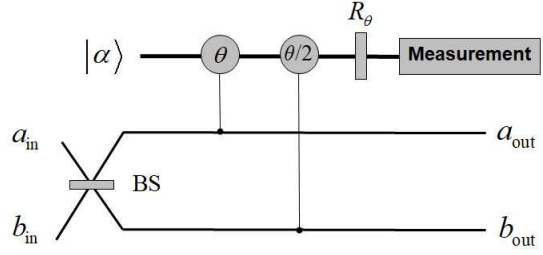}}
  \caption{The schematic diagram of symmetry detector for twin-beam six-photon entangled state by using linear optics and weak nonlinearities. $a_{\text{in}}$, $b_{\text{in}}$ are input ports of the 50:50 BS, and $a_{\text{out}}$, $b_{\text{out}}$ are the corresponding outputs, respectively. A coherent state $|\alpha\rangle$ is used for probe beam. $\theta$ and $\theta/2$ are phase shifts to characterize the interaction between photons in signal and probe modes. A single phase gate $R_{\theta}$ followed by an $X$ homodyne measurement provide an explicit result nearly deterministically.}
  \label{SD-3.eps}
\end{figure}

The nonlinear Kerr medium \cite{Imoto1985,SI1996,RV2005,Barrett2005} can be described intuitively as a medium which induce a cross phase modulation with Hamiltonian of the form
\(H = \hbar \chi a_s^\dag {a_s}a_p^\dag {a_p}\), where $a_s^\dag$ and ${a_s}$ ($a_p^\dag$ and ${a_p}$)
represent the creation and annihilation operators for the signal (probe), $\chi$ is the coupling strength of the nonlinearity.
This Kerr nonlinearity can be used to induce a phase modulation in probe mode depended on single-photon state and then it is capable of detecting the state of photons in signal mode nondestructively.
Since the Kerr nonlinearities are extremely weak \cite{Kok2008,LinHeBR2009,DYG2014,HDYG2015,HDYG2015OE}, we here restrict our discussion in the regime of weak nonlinearities.
The setup of symmetry detector is shown in Figure \ref{SD-3.eps}.

Consider a class of twin-beam six-photon entangled states
\begin{equation}\label{6-photon}
  \left|\psi\right\rangle= m(|3,0;0,3\rangle-|0,3;3,0\rangle) + n(|1,2;2,1\rangle-|2,1;1,2\rangle),
\end{equation}
where e.g. $|k,l;r,s\rangle$ as an abbreviation for state $|k\rangle_{a_{H}}\otimes|l\rangle_{a_{V}}\otimes|r\rangle_{b_{H}}\otimes|s\rangle_{b_{V}}$
means that there are $k$ horizontally polarized photons and $l$ vertically polarized photons in spatial mode $a$, and at the same time $r$ horizontally and $s$ vertically polarized photons in spatial mode $b$, and $m$ and $n$ are respectively the complex parameters satisfying the normalization condition $|m|^{2}+|n|^{2}=1/2$. Without loss of generality, we here suppose $m$ and $n$ are real and positive.

When the twin-beam six photons entangled state (\ref{6-photon}) passing through the 50:50 BS with exactly one beam per input-port, it yields
\begin{eqnarray}\label{6-photon-BS}
  \left|\psi'\right\rangle &=& \frac{1}{4}[(m+3n)(|3,0;0,3\rangle-|0,3;3,0\rangle) + (3m+n)(|1,2;2,1\rangle-|2,1;1,2\rangle)]  \nonumber \\
                      & &    + \frac{\sqrt{3}}{4}(m-n)(|3,2;0,1\rangle - |0,1;3,2\rangle + |1,0;2,3\rangle - |2,3;1,0\rangle).
\end{eqnarray}
Let $R_{\theta}=-9\theta/2$.
Due to the action of Kerr nonlinearities, the combined system evolves into
\begin{eqnarray}\label{6-photon-CS}
  \left|\psi'\right\rangle |\alpha\rangle &=&
         \frac{1}{4}[(m+3n)(|3,0;0,3\rangle-|0,3;3,0\rangle)|\alpha\rangle + (3m+n)(|1,2;2,1\rangle-|2,1;1,2\rangle)|\alpha\rangle]  \nonumber \\
& &    + \frac{\sqrt{3}}{4}(m-n)[(|3,2;0,1\rangle - |2,3;1,0\rangle)|\alpha \rm{e}^{i\theta}\rangle \nonumber \\
& &
+ (|1,0;2,3\rangle - |0,1;3,2\rangle)|\alpha \rm{e}^{-i\theta}\rangle ].
\end{eqnarray}

To proceed, we perform an $X$ homodyne measurement \cite{HDYG2015OE,Quantum-Noise2000} on the probe beam.
By means of the inner product $\langle x|\alpha\rangle=({2\pi })^{ - 1/4}\text{exp}[-(\text{Im}(\alpha))^2-(x-2\alpha)^2/4]$,
for $x > \alpha(1+\text{cos}\theta)$, we have
\begin{eqnarray}\label{6-photon-m1}
  \left|\psi_{0}\right\rangle &=& \frac{1}{\sqrt{10+24mn}}[(m+3n)(|3,0;0,3\rangle-|0,3;3,0\rangle) \nonumber \\
&&
  + (3m+n)(|1,2;2,1\rangle-|2,1;1,2\rangle)],
\end{eqnarray}
and for $x < \alpha(1+\text{cos}\theta)$, we have
\begin{eqnarray}\label{6-photon-m2-0}
  \left|\psi_{1}\right\rangle &=& \frac{1}{2}[\textrm{e}^{\textrm{i}\phi(x)}(|3,2;0,1\rangle - |2,3;1,0\rangle)
                                                + \textrm{e}^{-\textrm{i}\phi(x)}(|1,0;2,3\rangle - |0,1;3,2\rangle)],
\end{eqnarray}
where $\phi (x) = \alpha \sin \theta (x-2\alpha \cos \theta)\bmod 2\pi$, a phase shift related to the result of homodyne measurement.
Due to the overlaps between neighboring curves, the error probability of this measurement is given by $\varepsilon = { \textrm{erfc}\left( {{2\alpha(1-\rm{cos}\theta)}/2\sqrt 2 } \right)}/2$.
In one takes $\theta^2 \sim 10^{-2}$ and $\alpha \sim 10^{4}$, then $\varepsilon \sim 10^{-5}$, i.e., it is a nearly deterministic detection process in the regime of weak nonlinearities.

Note that the output state (\ref{6-photon-m1}) is exactly similar to the input state (\ref{6-photon}) and it will be dealt with using the similar procedures. We call these twin beams having the same number of photons per mode \textit{symmetric states}.
For the output state (\ref{6-photon-m2-0}), by applying a further phase shift operator $\textrm{e}^{\textrm{i}\phi(x)/2}$ in spatial mode $b$, for example, it yields
\begin{eqnarray}\label{6-photon-m2}
  \left|\psi'_{1}\right\rangle &=& \frac{1}{2}(|3,2;0,1\rangle - |0,1;3,2\rangle + |1,0;2,3\rangle - |2,3;1,0\rangle).
\end{eqnarray}
This state is different from those symmetric states because six photons are unequally shared between two spatial modes. It can also be collected and used to properly photonic quantum information processing.

For the outputting symmetric state, if one allows these photons to pass successively through symmetry detectors, where the input/output modes correspond to the signal photons, then it can tend towards six-photon entangled state $|\psi_3^{-}\rangle$.
To make this point more explicit, we rewrite $m$ and $n$ as $m_0$ and $n_0$, and let
\begin{eqnarray}\label{}
A=
\left(
  \begin{array}{cc}
    a_{00} & a_{01}\\
    a_{10} & a_{11}\\
  \end{array}
\right)
\end{eqnarray}
be a $2\times2$ matrix such that
\begin{eqnarray}\label{}
\left(
  \begin{array}{c}
    m_k\\
    n_k\\
  \end{array}
\right)
= A^{k}
\left(
  \begin{array}{c}
    m_0\\
    n_0\\
  \end{array}
\right).
\end{eqnarray}
Let $a_{00}=a_{11}=1$ and $a_{01}=a_{10}=3$.
Since $A$ is a Hermitian matrix, by taking
\begin{eqnarray}\label{}
T=
\frac {1}{\sqrt 2}\left(
  \begin{array}{cc}
    1 & 1\\
    1 & -1\\
  \end{array}
\right)
\end{eqnarray}
it can be diagonalized,
\begin{eqnarray}\label{}
\Lambda = T^{\dag}AT =
\left(
  \begin{array}{cc}
    4 & 0\\
    0 & -2\\
  \end{array}
\right).
\end{eqnarray}
Then we have
\begin{eqnarray}\label{}
A^{k}= T\Lambda^{k}T^{\dag}
     = \frac{1}{2}
\left(
  \begin{array}{cc}
    4^{k}+(-2)^{k} & 4^{k}-(-2)^{k} \\
    4^{k}-(-2)^{k} & 4^{k}+(-2)^{k} \\
  \end{array}
\right),
\end{eqnarray}
and thus
\begin{eqnarray}\label{}
\left(
  \begin{array}{c}
    m_k\\
    n_k\\
  \end{array}
\right)
     = \frac{1}{2}
\left(
  \begin{array}{cc}
     (4^{k}+(-2)^{k})m_{0} + (4^{k}-(-2)^{k})n_{0} \\
     (4^{k}-(-2)^{k})m_{0} + (4^{k}+(-2)^{k})n_{0} \\
  \end{array}
\right).
\end{eqnarray}
By this, after the $k$th iteration one can obtain the state
\begin{equation}\label{}
  \left|\psi\right\rangle= \frac{1}{\sqrt{C_{k}}}[m_{k}(|3,0;0,3\rangle-|0,3;3,0\rangle) + n_{k}(|1,2;2,1\rangle-|2,1;1,2\rangle)],
\end{equation}
where normalization factor $C_{k} = 2^{2k-1}(4^{k}+1) + 2^{2k+1}(4^{k}-1) m_{0}n_{0}$.

\begin{figure}[htbp]
\centerline{\includegraphics[width=3.2in]{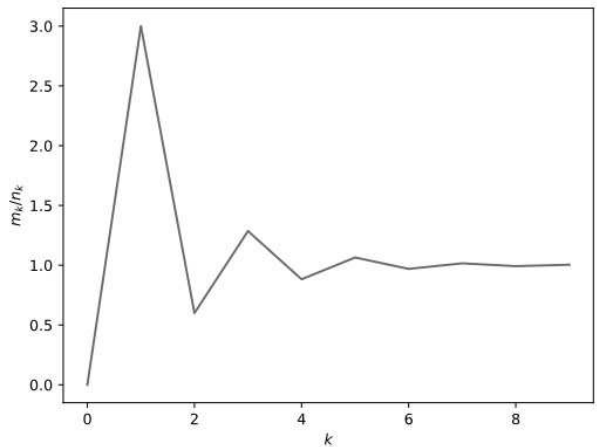}}
  \caption{The relationship between the ratio $m_{k}/n_{k}$ and the number of iterations $k$ ($k=1,2,\cdots,10$). As a result, with a few iterations the ratio will rapidly get close to $1$.}
  \label{iteration.eps}
\end{figure}

The ratio $m_{k}/n_{k}$ can be expressed as
\begin{eqnarray}\label{}
\frac{m_{k}}{n_{k}} =\frac{m_{0}+n_{0}+(-2)^{-k}(m_{0}-n_{0})}{m_{0}+n_{0}-(-2)^{-k}(m_{0}-n_{0})},
\end{eqnarray}
so for $k \rightarrow \infty$, $m_{k}/n_{k} = 1$.
Indeed, it is not difficult to see that with a few repeated iterations the ratio will approach $1$.
As an example, with initial value $m_{0}=0$, the relation between the ratios and the number of iterations is shown in Figure \ref{iteration.eps}.
Obviously, with 10 iterations the ratio gets rapidly close to $1$.
As a result, this provides an indirect method for purifying or preparing the six-photon entangled state $|\psi_3^{-}\rangle$ from the above twin-beam six-photon symmetric states.

\section{III. Preparation of multimode-photon entangled states}

The PDC photons are naturally entangled in energy and momentum, also can be manipulated and then entangled in polarization or path via photonic devices.
We now propose a scheme to create a class of six-photon entangled states in the form of polarization and path from third-order emission by using available linear optical elements.
Also, we describe a method to prepare six-photon polarization-entangled GHZ state by using weak nonlinearities.

\begin{figure}[htbp]
\centerline{\includegraphics[width=2in]{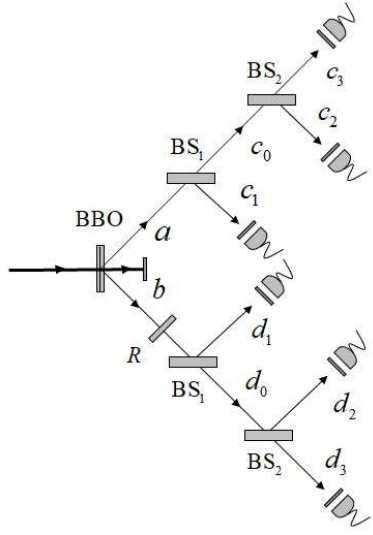}}
  \caption{The schematic diagram for creating six-photon entangled states from third-order down-conversion process.}
  \label{6-photon-he.eps}
\end{figure}

Consider the twin-beam six-photon entangled state
\begin{equation}\label{psi-3}
|\psi_3^{-}\rangle = \frac{1}{\sqrt{3!(3+1)!}}({a_{H}^{\dag}}{b_{V}^{\dag}} - {a_{V}^{\dag}}{b_{H}^{\dag}})^{3}|0\rangle,
\end{equation}
emitted from a femtosecond pump pulse acting on a type-II down-conversion crystal, BBO (beta barium borate).
A set of narrow-band filters are necessary to ensure that these down-converted photons are spectrally indistinguishable.
By design, this six-photon entangled state (\ref{psi-3}) will be in spatial modes $a$ and $b$, as shown schematically in Figure \ref{6-photon-he.eps}.

To proceed, a polarization rotation $R$, physically implemented by quarter- and half-wave plates, is used to rotate the polarization in spatial mode $b$ with the Hamiltonian $H_{R} = {i}\theta (b^{\dag}_{H} b_{V}-b_{H} b^{\dag}_{V})$.
Then, the six-photon state evolves into
\begin{eqnarray}\label{R}
\frac{1}{12} (\cos \theta {a_{H}^{\dag}}{b_{V}^{\dag}} - \cos \theta {a_{V}^{\dag}}{b_{H}^{\dag}} + \sin \theta {a_{H}^{\dag}}{b_{H}^{\dag}} + \sin \theta {a_{V}^{\dag}}{b_{V}^{\dag}})^{3}|0\rangle.
\end{eqnarray}
We next use two pairs of beam splitters $\rm{BS}_{1}$s and $\rm{BS}_{2}$s to split these photons into six spatial modes.
The reflection and transmission coefficients of each $\rm{BS}_{1}$ are respectively $R=1/3$ and $T=2/3$, with the evolution in operator form
\begin{eqnarray}\label{BS1}
a^{\dag} \rightarrow \sqrt{\frac{2}{3}}c^{\dag}_{0} + \sqrt{\frac{1}{3}}c^{\dag}_{1}, ~~~ b^{\dag} \rightarrow \sqrt{\frac{2}{3}}d^{\dag}_{0} + \sqrt{\frac{1}{3}}d^{\dag}_{1}.
\end{eqnarray}
Two $\rm{BS}_{2}$s are the most commonly used symmetric 50:50 beam splitters, characterized by the following transformation:
\begin{eqnarray}\label{BS2}
c^{\dag}_{0} \rightarrow \frac{1}{\sqrt{2}}(c^{\dag}_{2} + c^{\dag}_{3}), ~~~ d^{\dag}_{0} \rightarrow \frac{1}{\sqrt{2}}(d^{\dag}_{2} + d^{\dag}_{3}).
\end{eqnarray}
By this, the six-photon state (unnormalized) in spatial modes $c_{i}$ and $d_{i}$ ($i=1,2,3$) is described, in the operator form, by
\begin{eqnarray}\label{}
(\cos \theta c_{123,H}^{\dag}d_{123,V}^{\dag} - \cos \theta c_{123,V}^{\dag}d_{123,H}^{\dag} + \sin \theta c_{123,H}^{\dag}d_{123,H}^{\dag} + \sin \theta c_{123,V}^{\dag}d_{123,V}^{\dag})^{3} |0\rangle,
\end{eqnarray}
where, for simplicity, we write e.g. $c_{123,H}^{\dag}$ as an abbreviation for $c_{123,H}^{\dag}=c_{1H}^{\dag}+c_{2H}^{\dag}+c_{3H}^{\dag}$.

Due to the interference effects at the beam splitters, these six photons may be split into different spatial modes.
We herein consider the situation that there is only one photon in each of the spatial modes $c_{i}$ and $d_{i}$ ($i=1,2,3$).
By this, we arrive at an important class of six-photon entangled states
\begin{eqnarray}\label{polarization-path}
      & &|\Psi(\theta)\rangle\nonumber \\
 &=& \frac{1}{2}
\{
      [\cos^{3}\theta (|HHHVVV\rangle-|VVVHHH\rangle) \nonumber \\
&&
    + \sin^{3}\theta (|HHHHHH\rangle+|VVVVVV\rangle)]  \otimes   |c_{1}c_{2}c_{3}d_{1}d_{2}d_{3}\rangle  \nonumber \\
&&
   + \frac{1}{3} [\cos\theta(2\sin^{2}\theta-\cos^{2}\theta) (|HHVVVH\rangle-|VVHHHV\rangle) \nonumber \\
&&
                 +\sin\theta(\sin^{2}\theta-2\cos^{2}\theta)
                 (|HHVHHV\rangle+|VVHVVH\rangle)]  \nonumber \\
&&   \otimes     (|c_{1}c_{2}c_{3}\rangle + |c_{1}c_{3}c_{2}\rangle + |c_{2}c_{3}c_{1}\rangle)
                  (|d_{1}d_{2}d_{3}\rangle + |d_{1}d_{3}d_{2}\rangle + |d_{2}d_{3}d_{1}\rangle) \nonumber \\
&&
                 +\cos^{2}\theta \sin\theta [(|HHVVVV\rangle+|VVHHHH\rangle) (|c_{1}c_{2}c_{3}\rangle + |c_{1}c_{3}c_{2}\rangle + |c_{2}c_{3}c_{1}\rangle)
                  |d_{1}d_{2}d_{3}\rangle \nonumber \\
&&
                  +(|HHHVVH\rangle+|VVVHHV\rangle) |c_{1}c_{2}c_{3}\rangle (|d_{1}d_{2}d_{3}\rangle + |d_{1}d_{3}d_{2}\rangle + |d_{2}d_{3}d_{1}\rangle)
\nonumber \\  &&
                 +\cos\theta \sin^{2}\theta [(|HHHHHV\rangle-|VVVVVH\rangle) |c_{1}c_{2}c_{3}\rangle (|d_{1}d_{2}d_{3}\rangle + |d_{1}d_{3}d_{2}\rangle + |d_{2}d_{3}d_{1}\rangle)
\nonumber \\  && -(|HHVHHH\rangle-|VVHVVV\rangle) (|c_{1}c_{2}c_{3}\rangle + |c_{1}c_{3}c_{2}\rangle + |c_{2}c_{3}c_{1}\rangle)
                  |d_{1}d_{2}d_{3}\rangle]
\},
\end{eqnarray}
where, for convenience, we use path-state, e.g. $|c_{1}c_{2}c_{3}d_{1}d_{2}d_{3}\rangle$, to denote path information.
Note that a direct path operation can easily project the polarization-path entangled states onto polarization entangled states.
We take $\theta=\pi/2$ for example. With path-state $|c_{1}c_{2}c_{3}d_{1}d_{2}d_{3}\rangle$ one can write $|\Psi(\pi/2)\rangle$ as
\begin{eqnarray}\label{Psi-pi/2}
|\Psi(\pi/2)\rangle = \frac{1}{\sqrt{2}}|\rm{GHZ}_{6}^{}\rangle + \frac{1}{2}(|W_{3}\rangle|W_{3}\rangle+|\widetilde{W}_{3}\rangle|\widetilde{W}_{3}\rangle),
\end{eqnarray}
where
\begin{eqnarray}\label{}
|{\rm{GHZ}}_{6}\rangle = \frac{1}{\sqrt{2}}(|HHHHHH\rangle+|VVVVVV\rangle),
\end{eqnarray}
\begin{eqnarray}\label{}
|W_{3}\rangle = \frac{1}{\sqrt{3}}(|HHV\rangle+|HVH\rangle+|VHH\rangle),
\end{eqnarray}
and the ket $|\widetilde{W}_{3}\rangle$ is the spin-flipped $|W_{3}\rangle$.
In fact, the state $|\Psi(\pi/2)\rangle$ is equivalent to the state $|\Psi(0)\rangle$.
As an important application, such states can be used to exhibit perfect six-qubit correlations \cite{RWZB2009}.

\begin{figure}[htbp]
\centerline{\includegraphics[width=3.2 in]{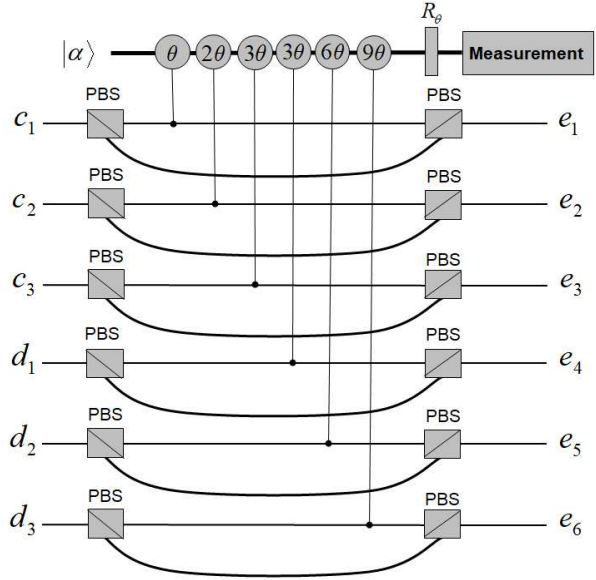}}
 \caption{The schematic diagram of preparing six-photon polarization-entangled GHZ state with linear optics and weak nonlinearities. $c_{i}$ and $d_{i}$ ($i=1,2,3$) are input ports related to the six signal modes, and $e_{i}$, $i=1,2,\cdots, 6$, are respectively the corresponding outputs. $\theta$, $2\theta$, $3\theta$s, $6\theta$ and $9\theta$ represent phase shifts in the coherent probe beam $|\alpha\rangle$ induced by Kerr media, respectively. $R_{\theta}$ is a single linear optical phase shifter.}
  \label{6-photon-ck.eps}
\end{figure}

Furthermore, we find that all such states in expression (\ref{polarization-path}) can be converted into six-photon polarization-entangled GHZ state.
It is possible to do this directly by using linear optics and weak nonlinearities.
We take the state $|\Psi(\pi/2)\rangle$ for example.
As show in Figure \ref{6-photon-ck.eps}, six input signal photons are individually split into two spatial modes at each PBS
and one mode of each photon will induce a phase shift on the coherent probe beam by using weak cross-Kerr media.
A further phase shift of $R_{\theta}=-12\theta$ in probe beam, achieved by a single linear optical phase shifter, is necessary to monitor overall phase.

Then the whole combined system evolves into
\begin{eqnarray}
|\Psi(\pi/2)\rangle_{\rm {ck}} & = &
  \frac{1}{2}(|HHHHHH\rangle|\alpha {\rm{e}}^{12\textrm{i}\theta}\rangle + |VVVVVV\rangle |\alpha {\rm{e}}^{-12\textrm{i}\theta}\rangle)
+\frac{1}{6}(|HHVHHV\rangle |\alpha\rangle   \nonumber \\
      & &
+ |VVHVVH\rangle |\alpha\rangle
            +|HVHHHV\rangle|\alpha {\rm{e}}^{\textrm{i}\theta}\rangle   + |VHVVVH\rangle |\alpha {\rm{e}}^{-\textrm{i}\theta}\rangle \nonumber \\
      & &
            +|VHHHHV\rangle|\alpha {\rm{e}}^{2\textrm{i}\theta}\rangle  + |HVVVVH\rangle |\alpha {\rm{e}}^{-2\textrm{i}\theta}\rangle
            +|HHVHVH\rangle|\alpha {\rm{e}}^{3\textrm{i}\theta}\rangle   \nonumber \\
      & &
            + |VVHVHV\rangle |\alpha {\rm{e}}^{-3\textrm{i}\theta}\rangle
            +|HVHHVH\rangle|\alpha {\rm{e}}^{4\textrm{i}\theta}\rangle  + |VHVVHV\rangle |\alpha {\rm{e}}^{-4\textrm{i}\theta}\rangle
 \nonumber \\
      & &
            +|VHHHVH\rangle|\alpha {\rm{e}}^{5\textrm{i}\theta}\rangle  + |HVVVHV\rangle |\alpha {\rm{e}}^{-5\textrm{i}\theta}\rangle
            +|HHVVHH\rangle|\alpha {\rm{e}}^{6\textrm{i}\theta}\rangle   \nonumber \\
      & &
            + |VVHHVV\rangle |\alpha {\rm{e}}^{-6\textrm{i}\theta}\rangle
            +|HVHVHH\rangle|\alpha {\rm{e}}^{7\textrm{i}\theta}\rangle  + |VHVHVV\rangle |\alpha {\rm{e}}^{-7\textrm{i}\theta}\rangle  \nonumber \\
      & &
            +|VHHVHH\rangle|\alpha {\rm{e}}^{8\textrm{i}\theta}\rangle  + |HVVHVV\rangle |\alpha {\rm{e}}^{-8\textrm{i}\theta}\rangle).
\end{eqnarray}
Consider the $X$ homodyne measurement on the probe beam. Then, by a similar argument, there exist ten intervals related to the value of the results of the homodyne measurement and each interval corresponds directly to an output state of the signal photons.

As a result, for $x < x_{0}=\alpha (\cos 12\theta + \cos 8\theta)$,
we have $1/\sqrt{2}(\textrm{e}^{\textrm{i}\phi_{12}(x)}|HHHHHH\rangle + \textrm{e}^{-\textrm{i}\phi_{12}(x)}|VVVVVV\rangle)$.
Let $x_{i}=\alpha [\cos (9-i)\theta + \cos (8-i)\theta]$ and $\phi_{i}(x)=\alpha [\cos (9-i)\theta + \cos (8-i)\theta]$, $i=1,2,\cdots,8$.
Then, for $x_{i-1} < x < x_{i}$,
we have $1/\sqrt{2}(\textrm{e}^{\textrm{i}\phi_{9-i}(x)}|HHHHHH\rangle + \textrm{e}^{-\textrm{i}\phi_{9-i}(x)}|VVVVVV\rangle)$,
up to two spin-flip transitions in the appropriate spatial modes.
For $x > x_{8}=\alpha (\cos \theta + 1)$, we have $1/\sqrt{2}(|HHVHHV\rangle + |VVHVVH\rangle)$.
At last, the undesired phase shifts $\phi_{i}(x)$, with $i=1,2,\cdots,8,12$, can be removed by placing a phase shifter in one of these modes depended on the value of the measurement.

\section{IV. Discussion and summary}

It is always permissible to prepare pairs of entangled photons by suppressing higher-order emissions.
However, a multiphoton coincidence may herald the higher-order down-conversion processes.
In order to characterize these photons emitted from PDC sources, in general, one takes $t_{d}$ and $t_{c}$ to be the pulse duration of the pump and the coherence time of the down-converted photons, respectively. Let $k = t_{d}/t_{c}$ be approximately the number of independent down-conversion processes in a pump pulse duration.
Consider a six-photon component which may include the first-, second-, and third-order processes.
There are three kinds of terms:
(\expandafter{\romannumeral 1}) the $k$ components
$|\psi_3^{-}\rangle \otimes |0\rangle \otimes \cdots \otimes |0\rangle + \cdots + |0\rangle \otimes \cdots \otimes |0\rangle \otimes |\psi_3^{-}\rangle$
where six photons are produced in a single third-order process;
(\expandafter{\romannumeral 2}) the $k(k-1)$ components
$|\psi_2^{-}\rangle \otimes |\psi_1^{-}\rangle \otimes |0\rangle \otimes \cdots \otimes |0\rangle + \cdots + |0\rangle \otimes \cdots \otimes |0\rangle \otimes |\psi_2^{-}\rangle \otimes |\psi_1^{-}\rangle$,
in which four photons are produced in a second-order process and one pair is produced in another first-order process;
(\expandafter{\romannumeral 3}) the $k(k-1)(k-2)/6$ components
$|\psi_1^{-}\rangle \otimes |\psi_1^{-}\rangle \otimes |\psi_1^{-}\rangle \otimes |0\rangle \otimes \cdots \otimes |0\rangle + \cdots + |0\rangle \otimes \cdots \otimes |0\rangle \otimes |\psi_1^{-}\rangle \otimes |\psi_1^{-}\rangle \otimes |\psi_1^{-}\rangle$
where three pairs are produced in different processes.
Then a normalized six-photon state can be expressed as
\begin{eqnarray}\label{}
|\Psi_{6}\rangle
 = \sqrt{\frac{6}{(k+1)(k+2)}} |\psi_3^{-}\rangle + \sqrt{\frac{6(k-1)}{(k+1)(k+2)}} |\psi_2^{-}\rangle\otimes|\psi_1^{-}\rangle + \sqrt{\frac{(k-1)(k-2)}{(k+1)(k+2)}} |\psi_1^{-}\rangle^{\otimes 3}.\nonumber\\
\end{eqnarray}

In this expression, it is interesting to note that when $k \ll 1$ then the main item is $|\psi_3^{-}\rangle$, while for $k \gg 1$ it remains nearly three pairs. In particular, for $k=1$, it only gives six-photon entangled state $|\psi_3^{-}\rangle$ because there is only one down-conversion process per pump pulse duration; for $k=2$, there may exist third-order emission or a mixture of first-order and second-order items since it is impossible to pump three-pair independent photons in two down-conversion processes.
Of course, experimentally, nowadays, not only entangled-photon pairs with high quality can be routinely created, but the twin-beam six-photon entangled state emitted from a single third-order PDC process can also be demonstrated by using a femtosecond pulse and narrow-band filters.

In summary, based on linear optics and weak nonlinearities we explicitly studied symmetry detection and multimode evolution for six-photon entangled state emitted from a single third-order PDC process.
Note that one can exactly pick out the right order process via suitable filtering and beam splitting \cite{ZZW1995}
and register a considerable six-fold coincidence with the aid of short pulses, narrow-band filters, and state-of-the-art single-photon detectors\cite{RWZB2009}.
In addition, we herein take weak nonlinearities into account, e.g. $\theta \sim 10^{-2}$, and thus our scheme becomes practically feasible in the current experimental techniques.
Due to the fact that there is no interferometric alignment to entangle independently emitted pairs, the strategy of preparing multimode polarization entangled states is strongly robust.
Since the present symmetry detector can be further extended to twin-beam $2n$-photon symmetric states, it is possible to provide a simple but novel approach to
produce entangled states involving a large number of photons by cascading a certain amount of symmetry detectors.

\section*{Acknowledgments}

We would like to thank Hai-Jun Tan for valuable discussions.
This work was supported by the National Natural Science Foundation of China under Grant Nos: 11475054, 11547169,
the Research Project of Science and Technology in Higher Education of Hebei Province of China under Grant Nos: QN2019305, Z2015188,
the Hebei Natural Science Foundation of China under Grant No:  A2018205125,
the Fundamental Research Funds for the Central Universities of Ministry of Education of China under Grant Nos: 3142019020, 3142017069.

\end{document}